\documentclass{article}
\usepackage{spconf,amsmath,graphicx}
\usepackage[inline]{enumitem}
\usepackage{hyperref}
\usepackage{url}
\usepackage{wrapfig}
\usepackage{caption}
\usepackage{float}
\title{Music Understanding LLaMA:\\Advancing Text-to-Music Generation\\with Question Answering and Captioning}

\name{Shansong Liu$^{1*}$, Atin Sakkeer Hussain$^{2*}$, Chenshuo Sun$^{2}$, Ying Shan$^{1}$
\thanks{$^*$Equal contribution. Shansong Liu is the corresponding author.}
}
\address{
    $^1$ ARC Lab, Tencent PCG \\
    $^2$ National University of Singapore
}

\begin{document}
%
\maketitle
\begin{abstract}
Text-to-music generation (T2M-Gen) faces a major obstacle due to the scarcity of large-scale publicly available music datasets with natural language captions. To address this, we propose the Music Understanding LLaMA (MU-LLaMA), capable of answering music-related questions and generating captions for music files. Our model utilizes audio representations from a pretrained MERT model to extract music features. However, obtaining a suitable dataset for training the MU-LLaMA model remains challenging, as existing publicly accessible audio question answering datasets lack the necessary depth for open-ended music question answering. To fill this gap, we present a methodology for generating question-answer pairs from existing audio captioning datasets and introduce the MusicQA Dataset designed for answering open-ended music-related questions. The experiments demonstrate that the proposed MU-LLaMA model, trained on our designed MusicQA dataset, achieves outstanding performance in both music question answering and music caption generation across various metrics, outperforming current state-of-the-art (SOTA) models in both fields and offering a promising advancement in the T2M-Gen research field.
\end{abstract}
\vspace{-0.2cm}

\section{Introduction}
\label{sec:intro}
\vspace{-0.2cm}

\noindent
For the task of music generation, the acquisition of substantial music data accompanied by captions is essential. However, most of the music datasets with accompanied captions are closed source data due to license restrictions \cite{copet2023simple, dhariwal2020jukebox, schneider2023mousai}. Presently, the largest publicly available dataset catering to this need is MusicCaps\cite{agostinelli2023musiclm}, comprising approximately 28.52 hours of music accompanied by captions. In comparison to other datasets available for tasks like audio classification or audio tagging, MusicCaps is relatively small. Therefore, there is an urgent requirement for developing a methodology that can generate text-music pairs on a large scale for public use.


Song Describer\footnote{\href{https://ismir2022program.ismir.net/lbd_405.html}{Song Describer}} is one of the few dedicated efforts to collect text-music pairs by crowd-sourcing for creating a public dataset. The authors built an online platform to recruit volunteers to annotate their provided music. Nonetheless, this approach is time-consuming and uncontrollable, and not suitable for obtaining large quantities of data for T2M-Gen research. As an alternative, we propose utilizing large language model (LLM) to automatically generate captions for vast amounts of music files from public resources.


Several models have been proposed for generating captions for music, including MusCaps \cite{Manco_2021}, audio captioning transformer \cite{mei2021audio}, audio captioning with audio-text retrieval pretraining \cite{xu2022sjtu}, Whisper tiny audio captioning \footnote{\href{https://huggingface.co/MU-NLPC/whisper-tiny-audio-captioning}{Whisper tiny audio captioning}}, and LP-MusicCaps \cite{doh2023lpmusiccaps}. Among these, MusCaps and LP-MusicCaps are currently the few specialized models dedicated explicitly to music captioning. The MusCaps model uses a hybrid architecture with a convolutional network for music understanding and recurrent neural network  for captioning. The LP-MusicCaps model uses a cross-modal encoder-decoder architecture to understand and caption music. An alternative approach for annotating music files involves employing audio question answering models to generate music captions.

Recently, several multi-model and LLM based models capable of audio understanding and question answering have emerged such as LLaMA Adapter \cite{zhang2023llamaadapter}, UniVAL \cite{shukor2023unified} and LTU \cite{gong2023listen}. LLaMA-adapter is a training scheme for finetuning LLMs. It offers multi-modal understanding capabilities based on ImageBind \cite{girdhar2023imagebind}. However, its ability to understand music is limited as it has not been trained on any music-text dataset. UniVAL was designed as a universal model for image, video, audio and language tasks. However, UniVAL also lacks pretraining with music-text data. The LTU model exhibits impressive performance and generalization abilities in audio question answering. However, it should be noted that the authors have not yet released the code and trained model or their constructed OpenAQA-5M dataset. Moreover, most of the training data are regular audio files rather than music, so it is still not an appropriate method for music question answering and music captioning.

In this paper, we present an innovative approach for generating text-music pairs to advance the field of text-to-music generation. To achieve this, we proposed a Music Understanding LLM built upon the Meta's LLaMA \cite{touvron2023llama, touvron2023llama2} model (MU-LLaMA) for music question answering. The proposed MU-LLaMA model\footnote{\href{https://github.com/crypto-code/MU-LLaMA}{MU-LLaMA model}} is capable of generating captions through answering music-related questions for the given music. In order to train MU-LLaMA, we designed and constructed a MusicQA dataset from two publicly available datesets, namely MusicCaps \cite{agostinelli2023musiclm} and MagnaTagATune \cite{9aed49b956a24e99b044582665fd5b21}.

This paper contributes significantly to both the domains of music question answering and text-to-music generation in the following noteworthy ways:
\begingroup
\renewcommand\labelenumi{ \textbf{\theenumi)}}
\begin{enumerate*}
    \setlength{\itemsep}{2mm}
    \item We introduce the MU-LLaMA model, an exceptional advancement capable of performing music question answering and captioning tasks, demonstrating superior performance across various metrics over SOTA models;

    \item We propose a systematic approach for creating the music question answering dataset, crucial for training the MU-LLaMA model;

    \item We demonstrate the use of the MU-LLaMA model to generate music captions in various formats required for developing T2M-Gen models.
\end{enumerate*}
\endgroup

This paper is organized as follows. In Section 2, we conduct a comprehensive comparison of different music representation methods to identify the most suitable music encoder for our proposed MU-LLaMA model. Section 3 outlines the methodology for creating the MusicQA dataset, crucial for training the MU-LLaMA model with the support of MosaicML's MPT model \cite{MosaicML2023Introducing}. Section 4 presents a detailed exposition of the MU-LLaMA model's structure and capabilities for music question answering and music captioning tasks. Experiments and evaluation of the MU-LLaMA model are done in Section 5. Finally, the conclusion summarizes key findings and contemplates potential future expansions.

\vspace{-0.2cm}
\section{Music Feature Representation}
\label{sec:music_rep}
\vspace{-0.2cm}

\noindent
In order to equip our MU-LLaMA model with music understanding capabilities, we employ pretrained audio representation models. These models can transform raw audio signals into meaningful representations that capture essential audio features, allowing machines to comprehend and interpret sound information. In this section, we compare the following audio representation models based on the performance of a music tagging task on the MagnaTagATune \cite{9aed49b956a24e99b044582665fd5b21} dataset.

\begin{table}[htbp]
\centering
\def\arraystretch{1.1}%
\caption{\textbf{Comparison of audio representation models.} (1) $MTT_{AUC}$: Accuracy of the predicted tags; (2) $MTT_{AP}$: Average precision of the predicted tags. The MERT model shows the bestperformance based on these metrics.}
\begin{tabular}{c|c|c}
\hline\hline
Model                       & \textbf{$MTT_{AUC}$}      & \textbf{$MTT_{AP}$}       \\ \hline\hline
\textit{\textbf{ImageBind}} & 88.55\%                & 40.19\%                \\
\textit{\textbf{JukeBox}}   & 91.5\%                 & 41.4\%                 \\
\textit{\textbf{OpenL3}}    & 89.35\%                & 42.88\%                \\
\textit{\textbf{CLAP}}      & 70.04\%                & 27.95\%                \\
\textit{\textbf{Wav2CLIP}}  & 90.15\%                & 49.12\%                \\
\textit{\textbf{MERT}}      & {\underline{\textbf{93.91\%}}} & {\underline{\textbf{59.57\%}}} \\ \hline\hline
\end{tabular}
\label{musictagging}
\end{table}

\begingroup
\renewcommand\labelenumi{ \textbf{\theenumi)}}
\begin{enumerate*}
    \item \textbf{ImageBind} \cite{girdhar2023imagebind} is a method that learns a joint embedding space across 6 modalities, including images, text, audio, depth, thermal and IMU data. It has the ability to represent and work with audio, enabling cross-modal retrieval and composition of modalities through arithmetic operations.

    \item \textbf{JukeBox} \cite{castellon2021codified} utilizes a multi-scale VQ-VAE and autoregressive Transformers to represent and generate music. Its music representation capabilities also extend to downstream tasks such as music information retrieval and music tagging.

    \item \textbf{OpenL3} \cite{8682475} is a framework for learning deep audio embeddings through unsupervised audio-visual correspondence, which can be used for various downstream tasks such as audio classification and analysis.

    \item \textbf{CLAP} \cite{wu2023largescale} leverages the publicly available LAION-Audio-630K dataset \cite{wu2023largescale} for contrastive language-audio pretraining, facilitating cross-modal audio-text retrieval.

    \item \textbf{Wav2CLIP} \cite{wu2022wav2clip} is constructed by distilling the CLIP \cite{radford2021learning} model, allowing audio to be projected into a shared embedding space together with images and text.

    \item \textbf{MERT} \cite{li2023mert}, an acoustic music representation model, achieves state-of-the-art performance on music understanding tasks through large-scale self-supervised training. It utilizes teacher models, including an acoustic teacher based on the RVQ-VAE model \cite{lai2022robust} and a musical teacher based on the constant-Q transform (CQT) \cite{Schrkhuber2010CONSTANTQTT} feature.
\end{enumerate*}
\endgroup

From Table \ref{musictagging}, MERT shows the best performance on the downstream task of music tagging on the MagnaTagATune (MTT) \cite{9aed49b956a24e99b044582665fd5b21} dataset and hence we choose the MERT model to generate music representation for our MU-LLaMA model.

\begin{figure*}
\centering
\includegraphics[width=\textwidth]{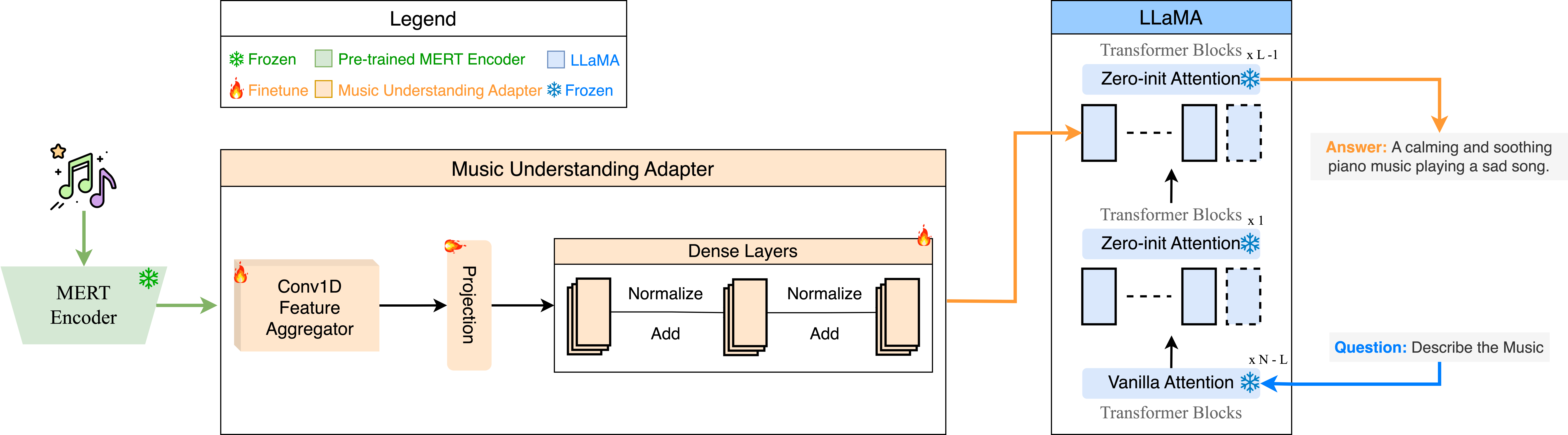}
\caption{\textbf{Music Understanding LLM (MU-LLaMA)}. The model is divided into 3 parts:
(1) Pretrained MERT Encoder to generate music representation;
(2) Music Understanding Adapter to fuse the representation into the LLaMA model;
(3) The LLaMA model that takes input from the adapter to learn music context information in the last $(L-1)$ layers.}
\label{mu-llama}
\end{figure*}

\vspace{-0.2cm}
\section{MusicQA Dataset Generation}
\label{sec:musicqa}
\vspace{-0.2cm}

\noindent
In order to equip our proposed MU-LLaMA model with music question answering capabilities, we require music question-answer pairs for training the model. Existing publicly available music datasets typically consist of descriptions or tags, lacking ready-made music question-answer pairs. Therefore, we propose an approach that leverages MosaicML's MPT-7B model \cite{MosaicML2023Introducing} to assist in generating music question-answer pairs. The MPT model can generate desired responses based on instructions. Therefore, we devise a set of instructions to generate music question-answer pairs from music captioning and music tagging datasets.

The first set of instructions guides the MPT model to generate answers based on the input caption or list of tags for the following questions:
\begingroup
\renewcommand\labelenumi{\textbf{\theenumi.}}
\begin{enumerate*}
    \setlength{\itemsep}{1mm}
    \item Describe the music;
    \item Describe the music in detail;
    \item What do you hear in the audio;
    \item What can be inferred from the audio.
\end{enumerate*}
\endgroup
The second set of instructions guides the MPT model to generate 5 open-ended question-answer pairs, which are related to music emotion, tempo, genre, etc., based on the input caption or a list of tags. Some possible generated questions are shown below:
\begingroup
\renewcommand\labelenumi{\textbf{\theenumi.}}
\begin{enumerate*}
    \setlength{\itemsep}{1mm}
    \item What is the mood of this audio;
    \item What instruments are being used in this audio;
    \item What is the tempo of this audio;
    \item What is the overall tone of this audio.
\end{enumerate*}
\endgroup
The MusicCaps\cite{agostinelli2023musiclm} and MagnaTagATune\cite{9aed49b956a24e99b044582665fd5b21} datasets are used to generate music question-answer pairs. MusicCaps contains music descriptions, so we utilize these descriptions to prompt the MPT model to generate question-answer pairs through paraphrasing. MagnaTagATune lacks descriptions but provides music tags. In this scenario, we leverage the inference ability of the MPT model to generate suitable question-answer pairs given the music tags. Due to space limitations, the instructions and some generated sample question-answer pairs will be shown in our demo page\footnote{\href{https://crypto-code.github.io/MU-LLaMA-Demo/}{MU-LLaMA demo page}}.

\vspace{-0.3cm}
\section{MU-LLaMA Model Architecture}
\label{sec:mu-llama}
\vspace{-0.2cm}

\noindent
Our MU-LLaMA model is built on Meta's LLaMA\cite{touvron2023llama,touvron2023llama2} model using MERT \cite{li2023mert} as the music encoder, which empowers the model for music understanding and question answering. To employ the MERT model, we use a similar approach as the LLaMA-Adapter \cite{zhang2023llamaadapter}. The architecture of our MU-LLaMA model is shown in Figure \ref{mu-llama}.

\begin{figure}[H]
\centering
\includegraphics[width=0.42\columnwidth]{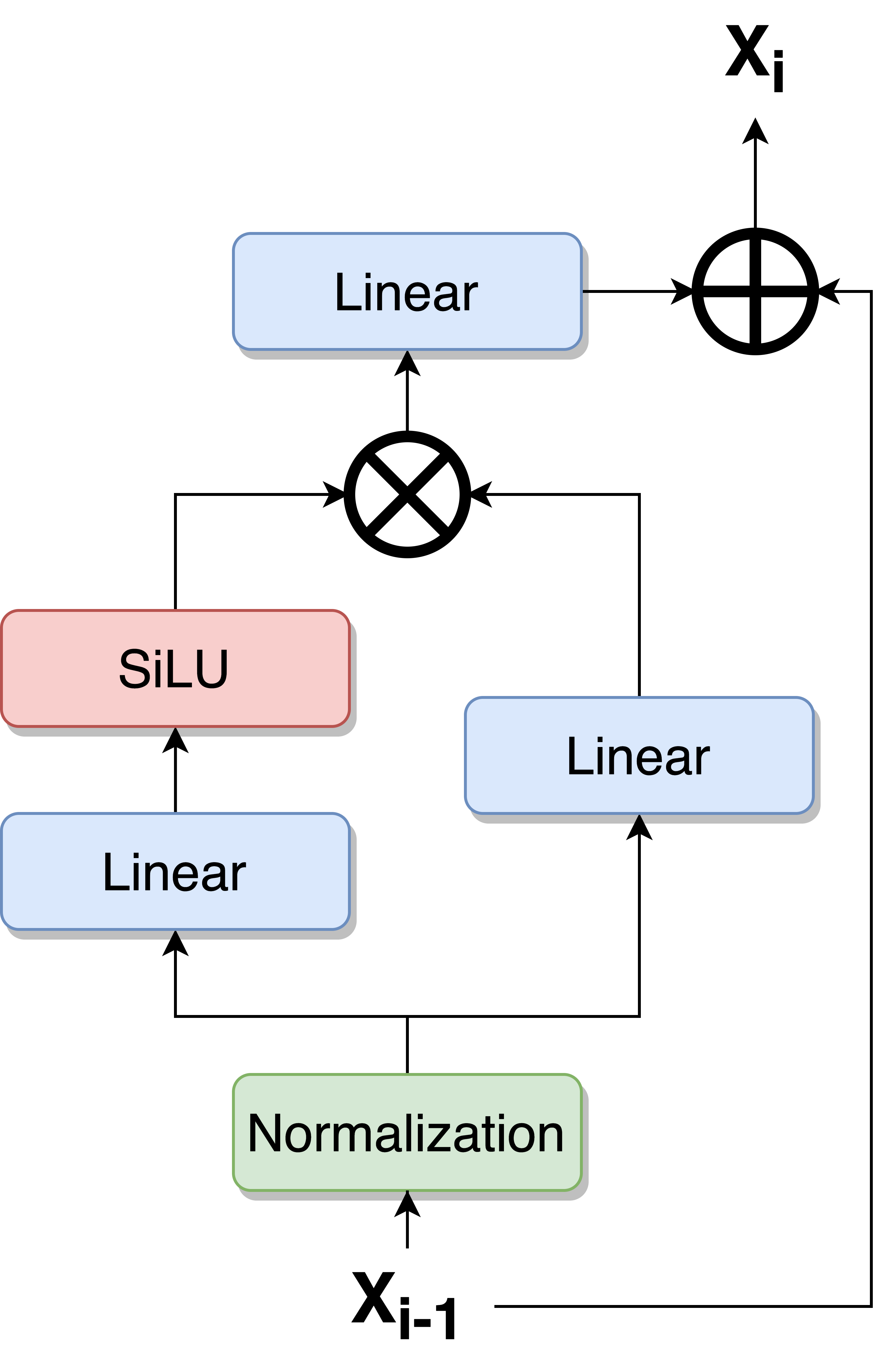}
\caption{\textbf{A Sub-block Structure of Music Understanding Adapter's Dense Layers}. Each sub-block is composed of normalization, linear layer, and SiLU activation function components. The input from the previous layer is also passed to the next layer through a skip connection.}
\label{mu-adapter}
\end{figure}

We use a frozen MERT encoder to generate music feature embeddings by stacking the outputs of the encoder's 24 hidden layers and 1 output layer. Each hidden layer and the output layer have a dimensionality of 1024. This results in a tensor of shape $(25, 1024)$. The subsequent 1D convolutional layer aggregates the feature embedding into a dimension of $1024$.  This is then projected to a $4096$ dimensional space by a projection layer and passed through a dense neural network containing 3 sub-blocks (see Figure \ref{mu-adapter}). Each sub-block consists of components such as normalization, linear layer, and Sigmoid Linear Unit (SiLU) activation function. The input from the previous layer is also passed to the next layer by a skip connection. This process is formulated as follows:
\vspace{-0.3cm}
\begin{multline*}
    X_{i} = X_{i-1} + L_{2,i}(SiLU(L_{1,i}(N_{i}(X_{i-1}))) \\ \times L_{3,i}(N_{i}(X_{i-1})))
\end{multline*}
\vspace{-0.5cm}

\noindent
where $X_0$ is the projected embedding output by the projection layer, $X_{i}$ is the embedding after the $i$-th sub-block of the dense neural network ($i\in\{1,2,3\}$), $L_{j,i}$ denotes the $j$-th linear layer in the $i$-th sub-block ($j\in\{1,2,3\}$), and $N_i$ denotes the normalization layer in the $i$-th sub-block. The final embedding $X_3$ with a feature dimension of $4096$ is used in the last $(L-1)$ layers of the LLaMA model to provide music context information in the question answering process.

\vspace{-0.2cm}
\vspace{-0.3cm}

During training, the MERT encoder \cite{li2023mert} and LLaMA's \cite{touvron2023llama} Transformer layers are frozen while the music understanding adapter is used for finetuning. The output from the adapter is multiplied to the queries in the multi-headed attention in the last $(L-1)$ transformer layers of the LLaMA model. Once the MU-LLaMA model is trained on our MusicQA dataset, it acquires the ability to answer questions given music context and can generate captions as well.

\vspace{-0.2cm}
\section{Experiments and Results}
\label{sec:experiments}


\subsection{Experiment Setup}

\noindent
\textbf{Dataset:} For the following experiments in this section, the music audios for evaluation are from the MTG-Jamendo dataset \cite{bogdanov2019mtg}, having no overlap with the music audios of our MusicQA dataset used for training the MU-LLaMA model. For the music question-answering subtask, we randomly selected 500 music tracks from the MTG dataset with at least 5 tags. Following the methodology used to create our MusicQA dataset, 9 music QA pairs are generated for each track, resulting in a total of 4500 QA pairs for evaluation, referred to as MTG-eval-QA. Regarding the music captioning subtask, we utilized the MPT model \cite{MosaicML2023Introducing} to generate music captions for 1000 randomly selected music tracks with at least 5 tags from the MTG dataset, referred to as MTG-eval-Cap. These generated music QA pairs or music captions serve as the references for model evaluation. The instructions used to generate this evaluation data will be showcased on our demo page.

\noindent
\textbf{Model Setup}: Our MU-LLaMA model utilizes the MERT model as the music encoder, followed by a dense neural network, producing 1024-dimensional music feature vectors. On the decoder side, we employ the LLaMA-2 7B model \cite{touvron2023llama2} and inject the music context information obtained from the MERT encoder into the top 19 layers $(L = 20)$ of the LLaMA-2 7B model. Our MU-LLaMA model was initially pretrained on the Alpaca Instruction-Following dataset and the MusicCaps subset of the MusicQA dataset, and then finetuned on the MTT subset of the MusicQA dataset. The base learning rate for both pretraining and finetuning was set to 0.0001 and batch size was set to 1 during the whole pretraining and finetuning processes. For the training epoch, the pretraining stage was set to 150 while the finetuning stage was set to 20. Gradient accumulation strategy was employed in both training stages to save GPU memory, achieving a larger effective batch size.

\noindent
\textbf{Evaluation Metric}: We evaluate the models using BLEU (B-U) \cite{papineni-etal-2002-bleu}, METEOR (M-R) \cite{banerjee-lavie-2005-meteor} ROUGE$_L$ (R-L) \cite{lin-2004-rouge} and BERT-Score (BERT-S) \cite{bert-score} which are common evaluation metrics for text generation. For the BLEU score, a weighted average of BLEU$_1$, BLEU$_2$, BLEU$_3$ and BLEU$_4$ ($weight = 0.25$ for each) is used.

\vspace{-0.2cm}
\subsection{Music Question Answering}
\label{ssec:musicqa_exp}

\noindent
We conducted experiments on the music question answering substask using a few available LLM based models capable of answering questions for input music, including LLaMA-Adapter \cite{zhang2023llamaadapter,gao2023llamaadapter} with ImageBind encoder and LTU model, as shown in Table \ref{musicqa_eval}. The results indicate that our MU-LLaMA model outperformed the other two compared models across all four metrics, particularly excelling in METEOR and ROUGE$_L$. It outperformed LTU by more than 10\% absolute in METEOR and ROUGE$_L$, while exceeded LLaMA Adapter by more than 5\% absolute in the same metrics.

\vspace{-0.3cm}
\subsection{Music Captioning}
\label{ssec:musiccap_exp}

\noindent
In the subtask of music captioning, we compared four models, including Whisper Audio Captioning (WAC), MusCaps, LTU and LP-MusicCaps, as shown in Table \ref{musiccap_eval}. The results demonstrate that our MU-LLaMA model remains the top performer in this subtask. We can observe that our MU-LLaMA model exhibits significant improvements over the four compared models in terms of BLEU, METEOR, and ROUGE$_L$ metrics. The elevated performance of the MU-LLaMA model can be attributed to the utilization of the pretrained MERT model for music comprehension and the finetuned LLaMA-2 model on our created MusicQA dataset for text generation.

\begin{table}[ht!]
\centering
\def\arraystretch{1.1}%
\caption{\textbf{Comparison of models for music question answering}. The best values of different metrics are made \textbf{bold}.}
\begin{tabular}{c|c|c|c|c}
\hline\hline
Model & \textbf{B-U$\uparrow$} & \textbf{M-R$\uparrow$} & \textbf{R-L$\uparrow$} & \textbf{BERT-S$\uparrow$} \\ \hline\hline
LTU & 0.242 & 0.274 & 0.326 & 0.887 \\
LLaMA Adapter & 0.273 & 0.334 & 0.413 & 0.895 \\
\textbf{MU-LLaMA} & \textbf{0.306} & \textbf{0.385} & \textbf{0.466} & \textbf{0.901} \\ \hline\hline
\end{tabular}
\label{musicqa_eval}
\end{table}
\vspace{-4mm}

\begin{table}[ht!]
\centering
\def\arraystretch{1.1}%
\caption{\textbf{Comparison of models for music captioning}. The best values of different metrics are made \textbf{bold}.}
\begin{tabular}{c|c|c|c|c}
\hline\hline
Model & \textbf{B-U$\uparrow$} & \textbf{M-R$\uparrow$} & \textbf{R-L$\uparrow$} & \textbf{BERT-S$\uparrow$} \\ \hline\hline
\textit{\textbf{WAC}} & 0.127 & 0.114 & 0.135 & 0.847 \\
\textit{\textbf{MusCaps}} & 0.121 & 0.125 & 0.140 & 0.849 \\
\textit{\textbf{LTU}} & 0.212 & 0.206 & 0.235 & 0.875 \\
\textit{\textbf{LP-MusicCaps}} & 0.227 & 0.177 & 0.227 & 0.877 \\
\textit{\textbf{MU-LLaMA}} & \textbf{0.278} & \textbf{0.261} & \textbf{0.312} & \textbf{0.888}\\ \hline\hline
\end{tabular}
\label{musiccap_eval}
\end{table}

\vspace{-0.5cm}
\section{Conclusion}
\label{sec:conclusion}

This paper introduces the MU-LLaMA model, designed for enhancing music-enabled question answering and captioning. Additionally, we present a methodology to construct music question answering datasets by leveraging existing music captioning and tagging datasets. The proposed MU-LLaMA model exhibits superior performance in both music question answering and music captioning tasks, surpassing the current state-of-the-art models while demonstrating strong generalization capabilities. Subsequent investigations are anticipated to center around the comprehension of speakers in the context of question answering related to the lyrical content in music. Additionally, there is potential to harness the MU-LLaMA model to curate datasets intended for text-to-music generation, thereby enhancing the efficacy of T2M-Gen models.

\vfill\pagebreak


\bibliographystyle{IEEEbib}

{\small
\bibliography{refs}}

\end{document}